\pdfoutput=1
\documentclass[3p,times,procedia]{elsarticle}
\usepackage{nupha_ecrc}
\usepackage{amsmath}
\usepackage{graphicx}
\usepackage{caption}
\usepackage{subcaption}
\usepackage{placeins}
\usepackage{lineno}
\usepackage{amssymb}
\usepackage[figuresright]{rotating}
\usepackage{xspace}

\volume{00}

\firstpage{1}

\journalname{Nuclear Physics A}

\runauth{}

\jid{nupha}

\jnltitlelogo{Nuclear Physics A}

\newcommand{\nudyn}{$\nu_{dyn}$\xspace}
\newcommand{\nudynpika}{$\nu_{dyn}[\pi^{+}+\pi^{-},K^{+}+K^{-}]$\xspace}
\newcommand{\nudynpipr}{$\nu_{dyn}[\pi^{+}+\pi^{-},p+\overline{p}]$\xspace}
\newcommand{\nudynkapr}{$\nu_{dyn}[K^{+}+K^{-},p+\overline{p}]$\xspace}
\newcommand{\pt}{\ensuremath{p_{\rm T}}\xspace}
\newcommand{\dEdx}{d$E$/d$x$\xspace}

\newcommand{\dNdeta}{d$N_{\rm ch}/$d$\eta$\xspace}

\newcommand{\snn}{$\sqrt{s_{NN}}$\xspace}
\sloppypar

\begin{document}

\begin{frontmatter}

\dochead{}
\title{Event-by-Event Identified Particle Ratio Fluctuations in Pb-Pb Collisions with ALICE using the Identity Method}
\author[1]{Mesut Arslandok (for the ALICE Collaboration)}
\address[1]{Institut f\"{u}r Kernphysik, Goethe Universit\"{a}t Frankfurt am Main, Germany}

\begin{abstract}
\indent The study of event-by-event fluctuations of identified hadrons may reveal the degrees of freedom of the strongly 
interacting matter created in heavy-ion collisions and the underlying dynamics of the system. 
The observable \nudyn, which is defined in terms of the moments of identified-particle multiplicity 
distributions, is used to quantify the magnitude of the dynamical fluctuations in event-by-event 
measurements of particle ratios. The ALICE detector at the LHC is well-suited for the study of 
\nudyn, due to its excellent particle identification capabilities. Particle identification 
based on the measurement of the specific ionisation energy loss, \dEdx, works well on a statistical basis 
but suffers from ambiguities when applied on an event-by-event level.  
A novel experimental technique called the "Identity Method" is used to overcome such limitations. 
The first results on identified particle ratio fluctuations in Pb-Pb 
collisions at \mbox{\snn=2.76~TeV} in ALICE as a function of centrality are presented. 
The ALICE results for the most peripheral events indicate an increasing correlation between pions and protons 
which is not reproduced
by the HIJING and AMPT models. On the other hand, for the most central events the ALICE results 
agree with the extrapolations based on the data at lower energies from CERN-SPS and RHIC.
\end{abstract}


\end{frontmatter}


\FloatBarrier
\section{Introduction}
\label{}

The theory of strong interaction, Quantum Chromodynamics (QCD), predicts that at sufficiently high energies 
nuclear matter transforms into a deconfined state where nucleons are no longer the basic constituents but rather 
quarks and gluons. The event-by-event fluctuations \cite{shuryak} of suitably chosen observables opens new possibilities to 
investigate the properties of the so called "QCD phase diagram".

In general, one can classify fluctuations into two parts. First, dynamical fluctuations which result from correlated particle 
production reflecting the underlying dynamics of the system. Second, statistical fluctuations induced by the measurement process 
itself due to the finite event multiplicity. 

In case of independent particle production the multiplicity fluctuations follow a Poisson distribution. 
To quantify particle correlations the observable \nudyn was proposed in Ref.~\cite{voloshin, voloshin2}. It is defined as

\begin{equation} \label{eq:nudyn}
\nu_{dyn}[A,B] = \dfrac{\langle N_{A}^{2} \rangle}{{\langle N_{A} \rangle}^{2}} + 
\dfrac{\langle N_{B}^{2} \rangle}{{\langle N_{B} \rangle}^{2}} - 
2\dfrac{\langle N_{A}N_{B} \rangle}{\langle N_{A} \rangle\langle N_{B} \rangle} -
\left(\dfrac{1}{\langle N_{A} \rangle} +\dfrac{1}{\langle N_{B} \rangle}\right),
\end{equation} 

\noindent where $N_{A}$ and $N_{B}$ are the numbers of particle types A and B in
a given event and $\langle...\rangle$ denotes averaging over events. The quantity \nudyn vanishes when the multiplicity 
distributions of both particle types are Poissonian and independent from each other. 
Since the negative terms in the formula of \nudyn contain a correlation term, negative values of \nudyn 
indicate the degree of correlation between particle species.

In this article, we present the measurements of \mbox{\nudynpika}, \mbox{\nudynpipr} and \mbox{\nudynkapr} in Pb-Pb 
collisions at \snn=2.76~TeV in ALICE. The results are compared to the HIJING \cite{HIJINGref} and AMPT \cite{AMPTref} models 
as well as to measurements at lower energies from CERN-SPS \cite{NA49} and RHIC \cite{STAR}.

\FloatBarrier
\section{Analysis Details}

The present analysis uses about 13 million minimum bias Pb-Pb events at \snn=2.76~TeV collected by the ALICE detector \cite{ALICE} 
during the 2010 LHC run. The subdetectors involved in the analysis were the Time Projection Chamber (TPC) 
for the tracking and particle identification and the Inner Tracking System (ITS) for the vertexing. Moreover, the two forward 
V0 detectors were used for the centrality estimation. Particle identification (PID)
was based on the measurement of the specific ionisation energy loss, \dEdx, in the TPC gas. The charged particles 
reconstructed in the TPC in full azimuth and in the pseudo-rapidity range of $|\eta|<$0.8 were used in this analysis. 
The momentum range was restricted to 0.2$<$$p$$<$1.5~GeV/$c$ in order to minimise the systematic uncertainties due to overlap of 
measured particle dE/dx distributions.   

\FloatBarrier
\subsection{Identity Method}

When the \dEdx distributions of different particle species are well separated, a unique particle identification is possible.
However, this is not possible when the measured particle \dEdx distributions overlap.

\begin{figure}[h]
 \centering
   \includegraphics[width=11cm]{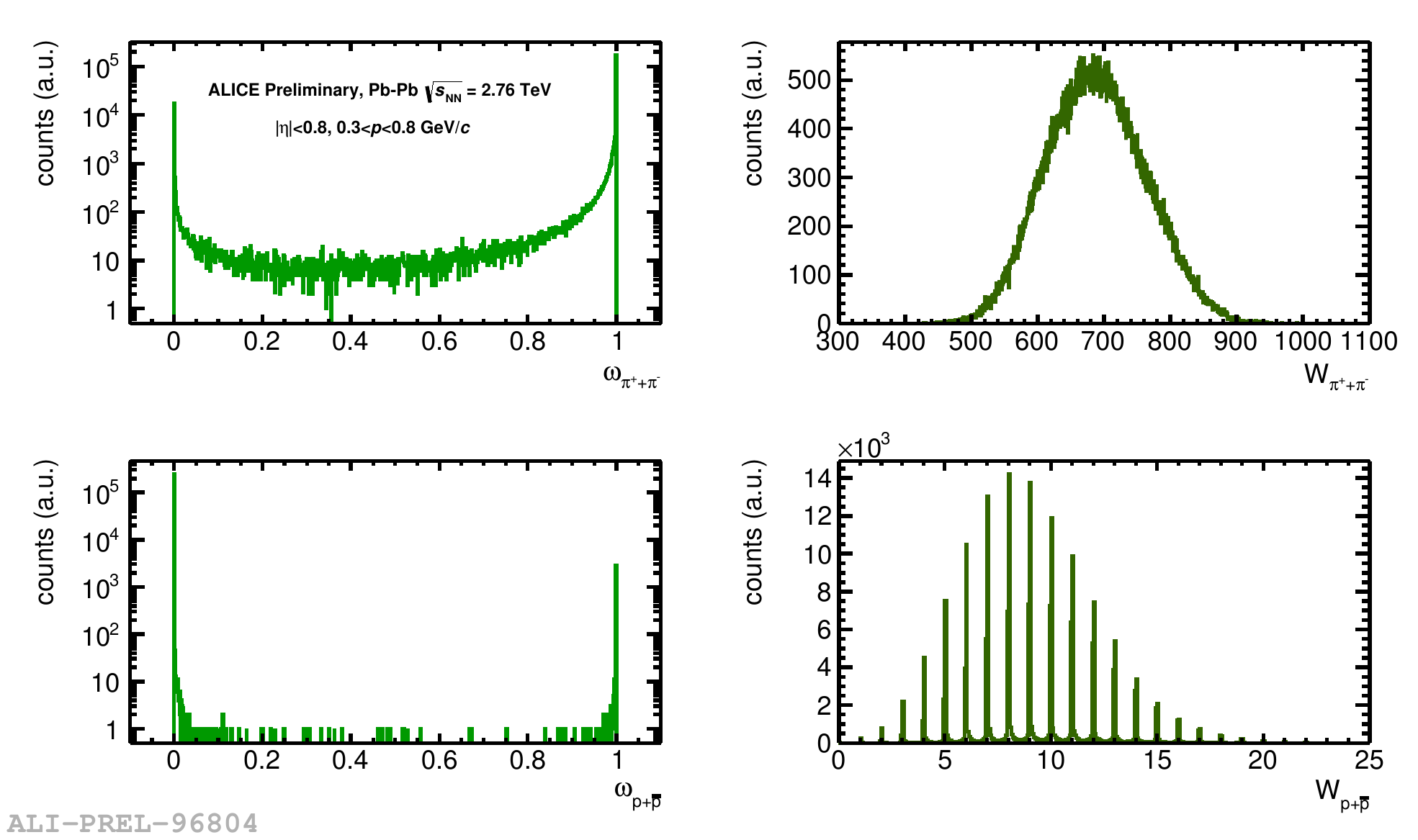}
 \caption{ $\omega$ and $W$ distributions for pions and protons within the momentum range of 0.3$<$$p$$<$0.8~GeV/$c$.}
\label{fig:wdist}
\end{figure}

The so called "Identity Method" \cite{marek} proposes a solution to this misidentification problem. 
The method follows a probabilistic approach using the inclusive \dEdx distributions measured in the TPC, and determines 
the moments of the multiplicity distributions by an unfolding procedure. It requires two basic 
experimentally measurable event-by-event quantities, $\omega$ and $W$, which are defined as

\begin{equation} \label{eq:omega}
\omega_{j}(x) = \dfrac{\rho_{j}(x)}{\rho(x)} \longrightarrow \omega_{j}\in[0,1], \quad \quad  W_{j} \equiv \sum_{i=1}^{N(n)} \omega_{j}(x_{i}),
\end{equation}

\noindent where $x$ stands for the \dEdx of a given track, 
$\rho_{j}(x)$ is the \dEdx distribution of particle $j$, $\rho(x)$ is the inclusive \dEdx 
distribution within a given event and $N(n)$ is the number of tracks within the $n$th event. In other words,
$\omega_{j}$ is a Bayesian probability measure of being particle type $j$ for a given track. Thus, in case of perfect 
identification one expects $\langle W_{i} \rangle = \langle N_{i} \rangle$, while this does not hold in case of 
misidentification. 

The moments of the $W$ distributions $\langle W_{i} ... W_{j} \rangle$ can be easily constructed 
directly from experimental data. The identity method provides the real moments $\langle N_{i} ... N_{j} \rangle$ 
of the particle multiplicity distributions by means of an unfolding procedure using the moments of these $W$ 
distributions. Examples for distributions of $\omega$ and $W$ in Pb-Pb data are shown in Fig.~\ref{fig:wdist}. 
Details of the derivation can be found in Ref.~\cite{goren, proofTiden}. 

\FloatBarrier
\subsection{Particle Identification (PID)}

\begin{figure}[h]
\centering
\hspace*{-0.4cm} 
 \begin{tabular}[c]{cccc}
 
  \begin{minipage}[b]{0.5\textwidth}
     \centering
     \includegraphics[width=8cm]{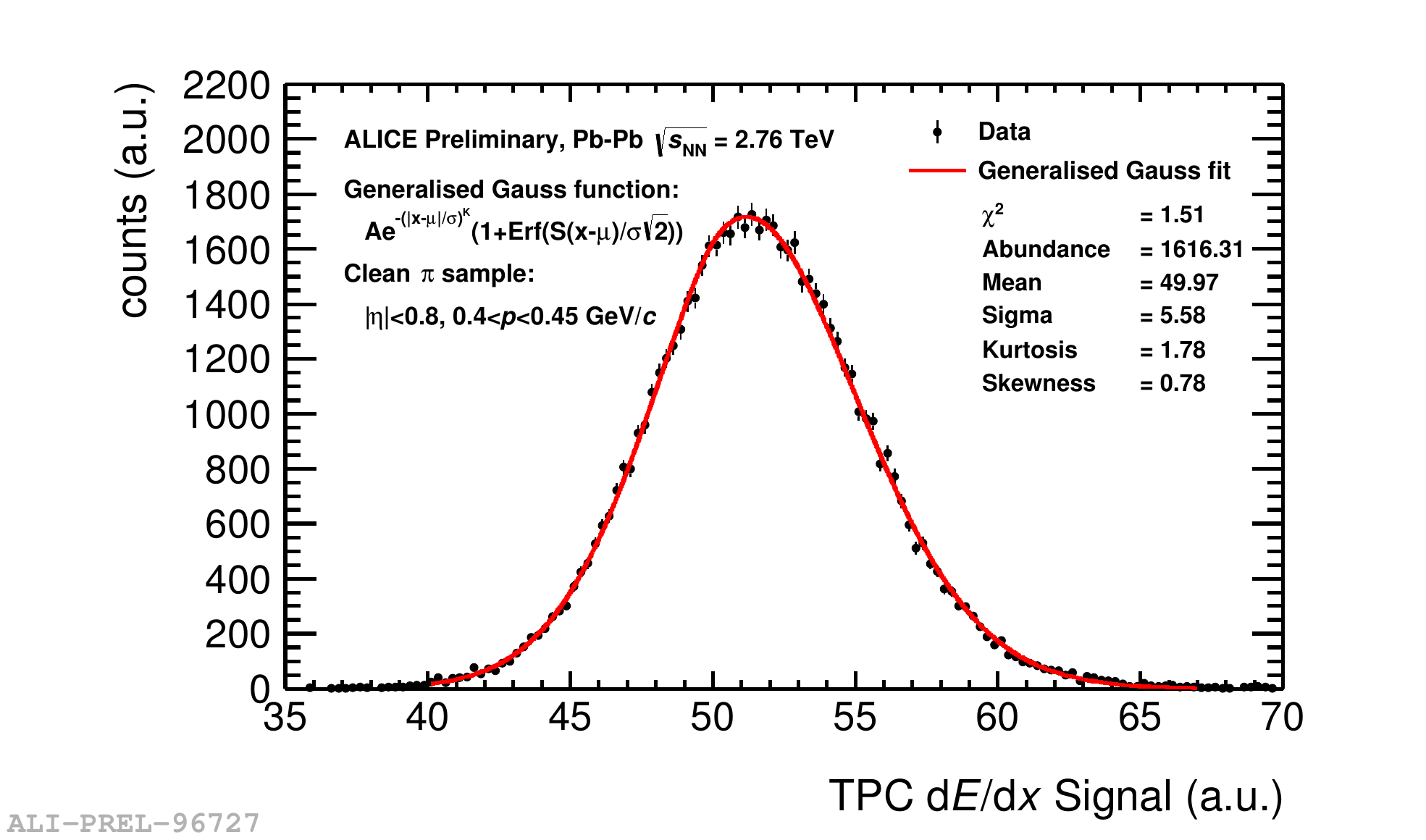}
  \end{minipage}&
  
  \begin{minipage}[b]{0.5\textwidth}
     \centering
     \includegraphics[width=8cm]{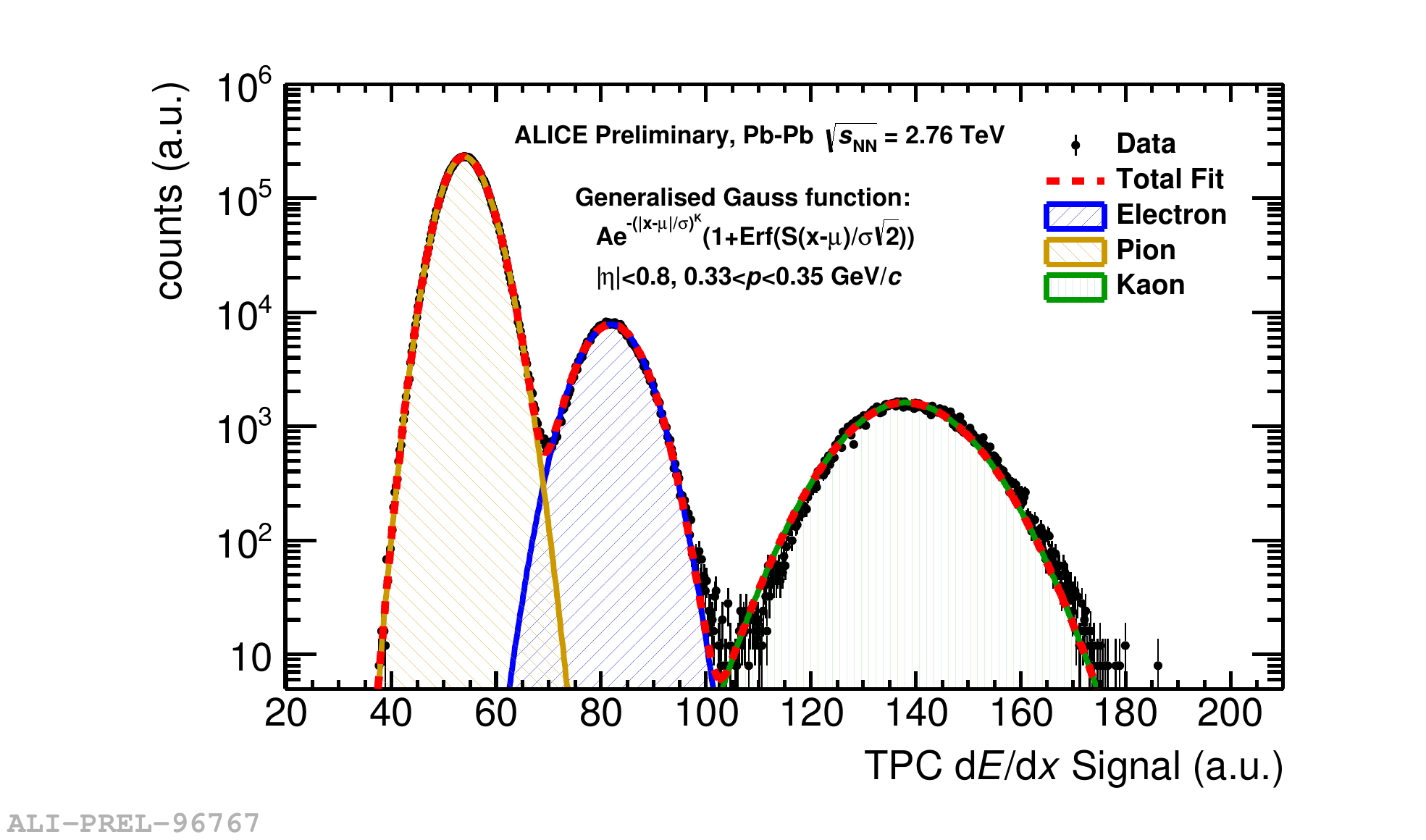}
  \end{minipage}&
 
 \end{tabular}
\caption{Left: \dEdx distribution of a clean pion sample retrieved from the decay of $K_{S}^{0}$. A generalised 
Gauss function was used for the fit. Right: \dEdx distributions of pions, kaons and electrons within the momentum interval of
0.33$<$$p$$<$0.35~GeV/$c$.}
\label{fig:dedxdist}
\end{figure}

The identity method exploits the fits of inclusive \dEdx distributions for the calculation of $\omega$. Since the overlap 
regions in the \dEdx distributions are also allowed, one needs very good fits of inclusive \dEdx spectra over the full 
momentum range covered in the analysis. Therefore, a good understanding of the detector response of the TPC is required. 
To this end, clean particle samples, which were retrieved from V0 decay particles (pions from $K_{S}^{0}$ decays, protons from 
$\varLambda$ decays and electrons from photon conversions), were used. Fig.~\ref{fig:dedxdist} left panel shows the determination 
of the detector response for pions, where a generalised Gauss function was used for a better description of the response shape. 
The detector response functions obtained in this way were later used as the fit functions in the inclusive 
\dEdx spectra fits as shown in Fig.~\ref{fig:dedxdist} right panel.

\FloatBarrier
\section{Results}

\begin{figure}[h]
\centering
\hspace*{-1cm} 
 \begin{tabular}[c]{cccc}
  \begin{minipage}[b]{0.33\textwidth}
     \centering
     \includegraphics[width=5.9cm]{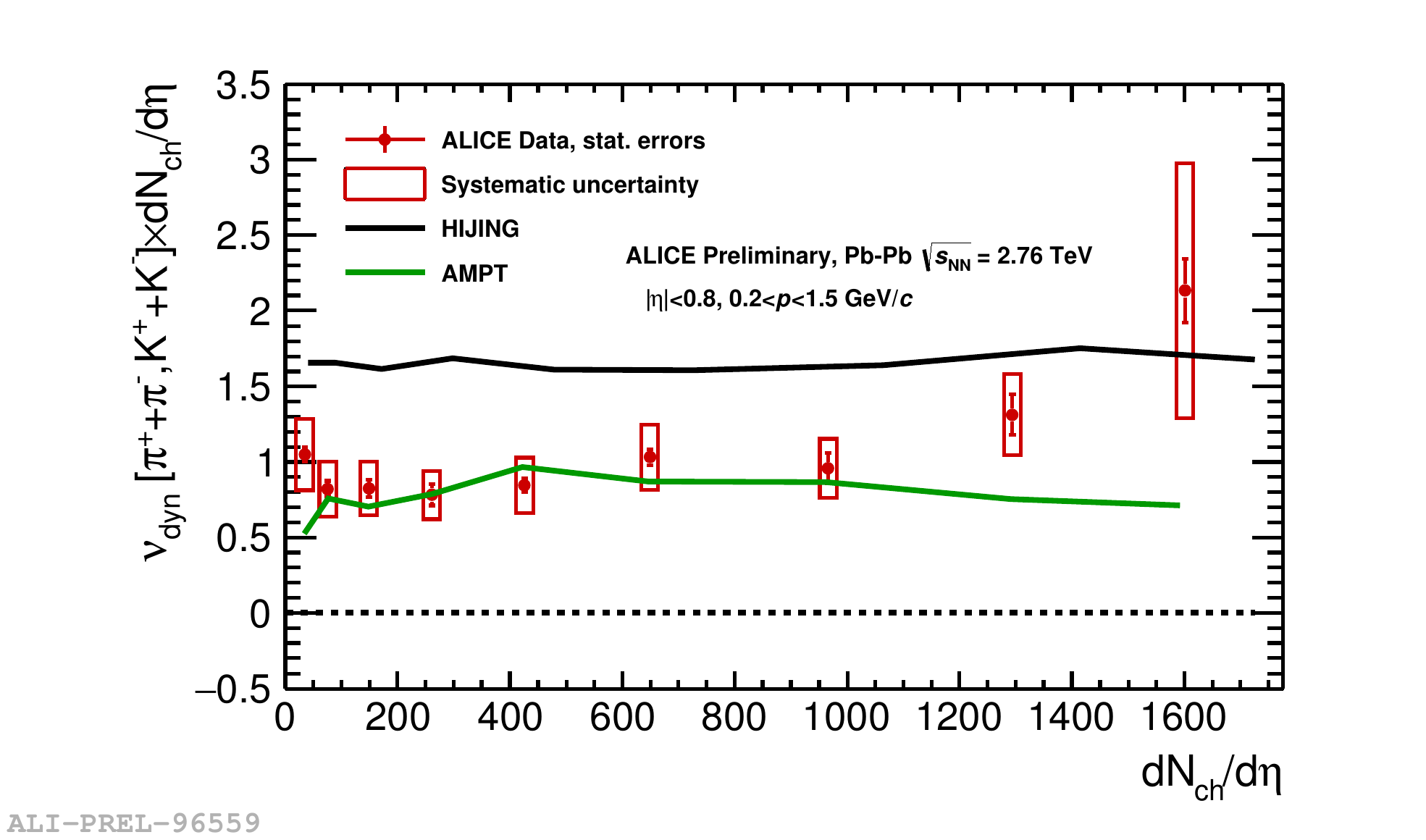}
  \end{minipage}&
  
  \begin{minipage}[b]{0.33\textwidth}
     \centering
     \includegraphics[width=5.9cm]{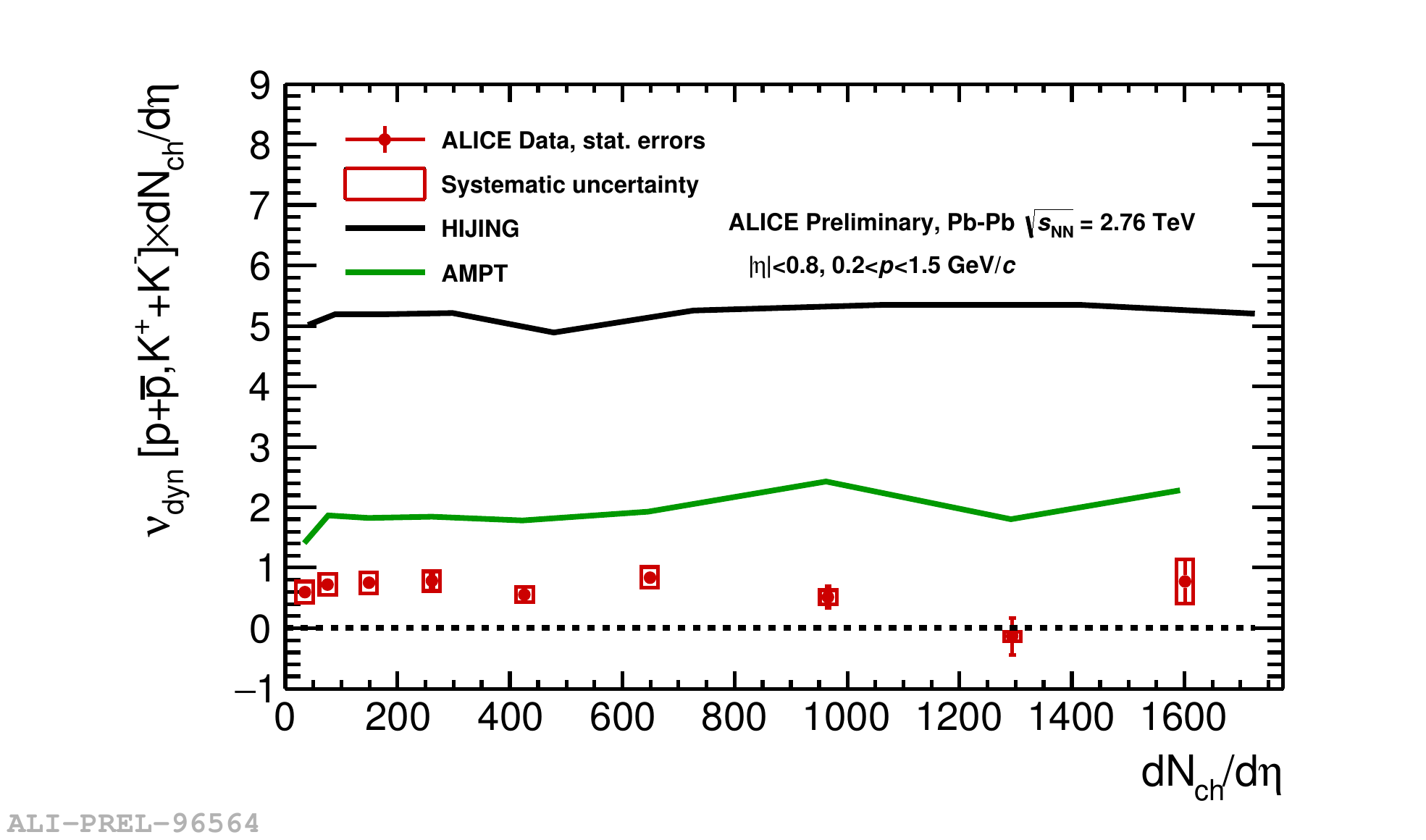}
  \end{minipage}&
  
  \begin{minipage}[b]{0.33\textwidth}
     \centering
     \includegraphics[width=5.9cm]{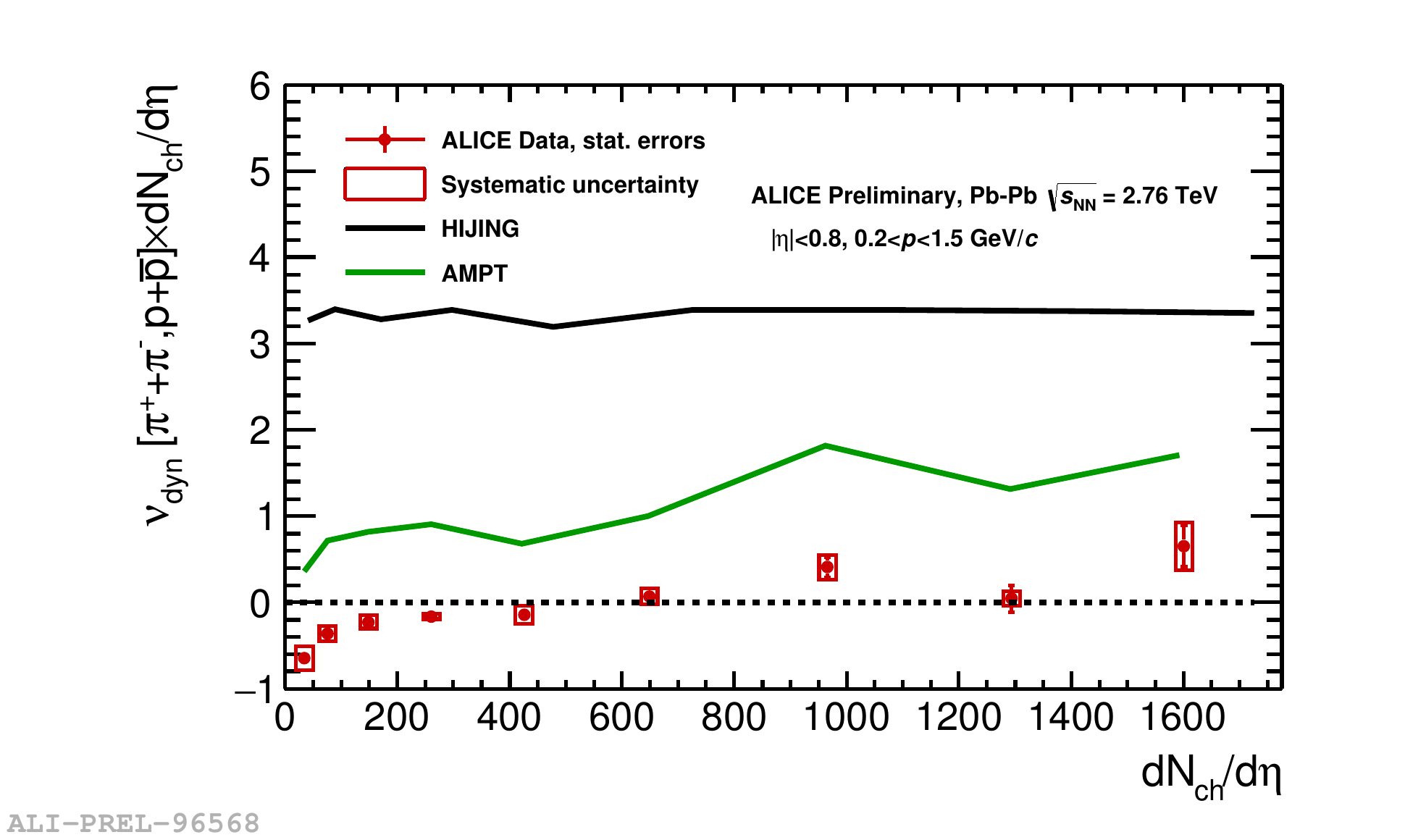}
  \end{minipage}&
 \end{tabular}
\caption{ALICE results for \nudyn scaled by \dNdeta. Black and green solid lines show HIJING \cite{HIJINGref} and AMPT
 \cite{AMPTref} model calculations, respectively.}
\label{fig:centralitydep}
\end{figure}

The results on particle ratio fluctuations are presented in terms of \nudyn, which is constructed from the 
moments of the $W_{j}$ distributions (Eq.~\ref{eq:nudyn}, \ref{eq:omega}). We have verified with a Monte Carlo simulation
that the finite reconstruction 
efficiency affects the results for \nudyn by less than 10\%, which was added to the systematic uncertainties. 
The other two main contributions to the systematic uncertainties are from the inclusive \dEdx spectra fits (15$\%$) and 
the cut on the maximum distance of the reconstructed vertex to the nominal interaction point along the beam axis (5$\%$).
 Fig.~\ref{fig:centralitydep} shows results for \mbox{\nudynpika}, \mbox{\nudynpipr} and \mbox{\nudynkapr} divided by 
 the charged-particle 
 multiplicity density at mid-rapidity, \dNdeta, as a function of the collision centrality, also expressed in terms of \dNdeta \cite{spectra}. 
 The results are compared to calculations with the HIJING \cite{HIJINGref} and AMPT \cite{AMPTref} event generators. In this representation, the HIJING results are almost 
independent of centrality, reflecting that HIJING is based on a superposition of independent p-p collisions. 
AMPT, which includes initial state fluctuations and collectivity, shows a weak centrality dependence. Both models 
exceed the data with the exception of \nudynpika, where AMPT is in reasonable agreement with the data.

\begin{figure}[h]
\centering
\hspace*{-1cm} 
 \begin{tabular}[c]{cccc}
  \begin{minipage}[b]{0.33\textwidth}
     \centering
     \includegraphics[width=5.9cm]{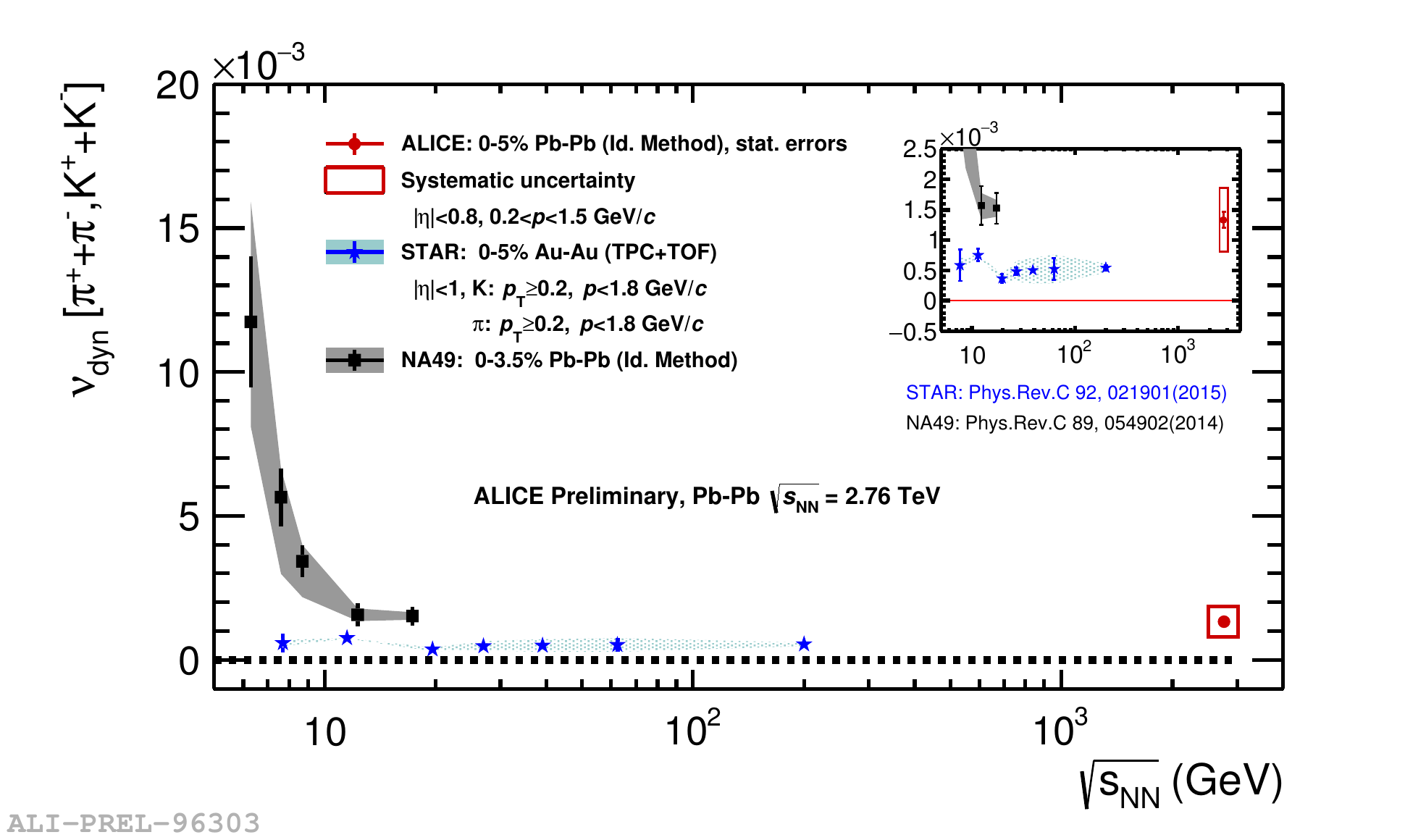}
  \end{minipage}&
  
  \begin{minipage}[b]{0.33\textwidth}
     \centering
     \includegraphics[width=5.9cm]{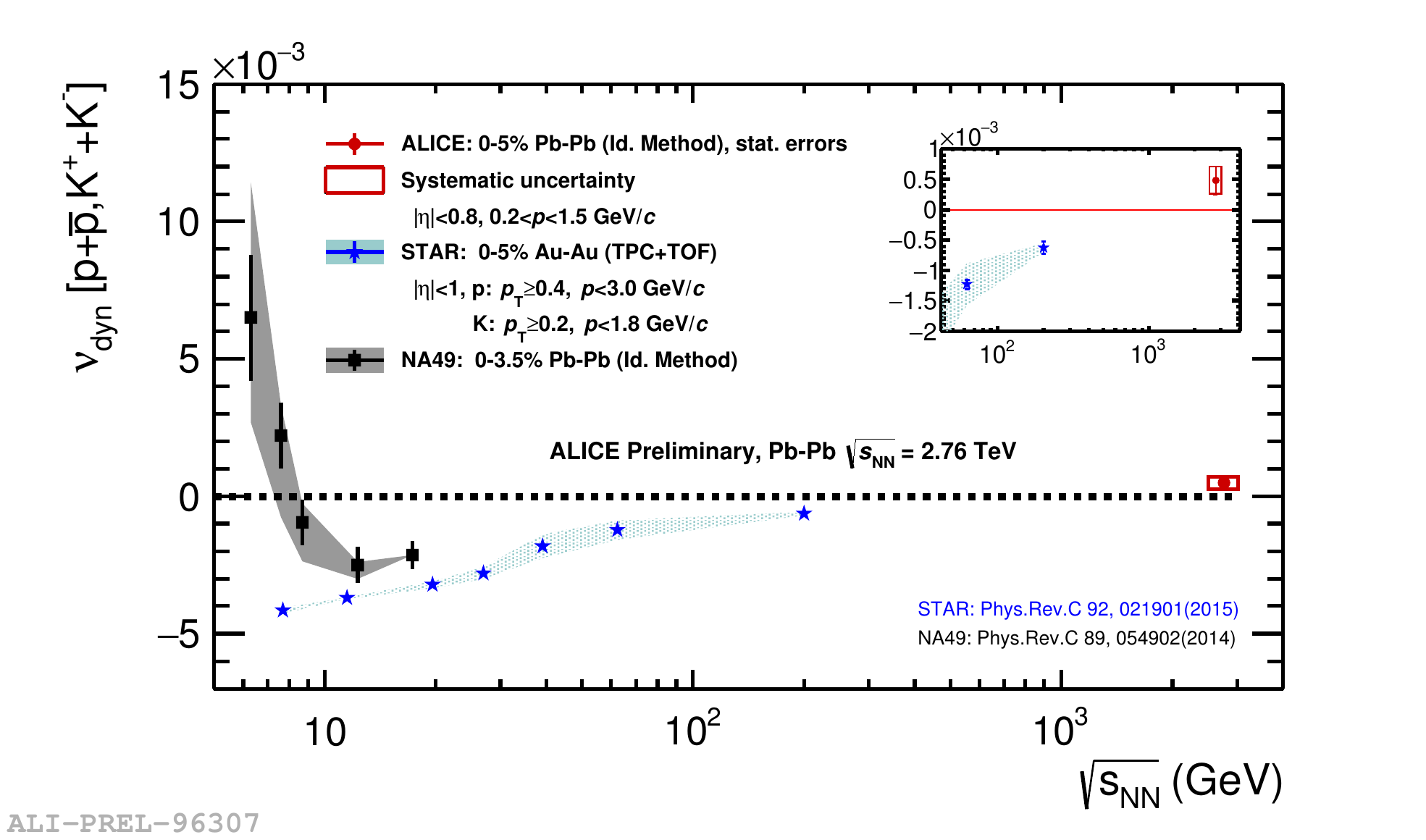}
  \end{minipage}&
  
  \begin{minipage}[b]{0.33\textwidth}
     \centering
     \includegraphics[width=5.9cm]{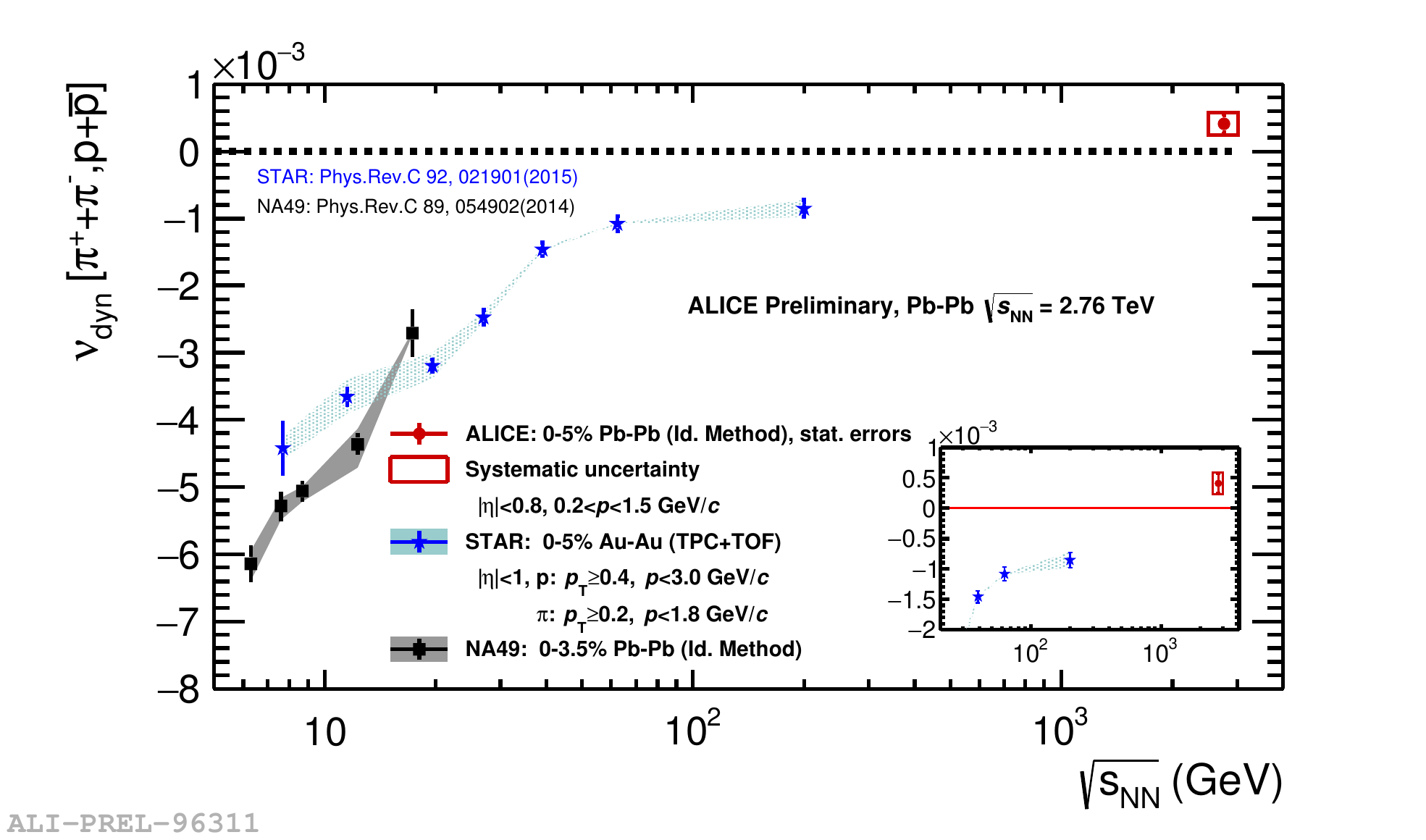}
  \end{minipage}&
 \end{tabular}
\caption{Energy dependence of $\nu_{dyn}$. ALICE results are shown as red solid circles. Results from the Identity method 
for central Pb-Pb collisions from NA49 \cite{NA49} are shown with black solid squares. Stars represent results from STAR 
\cite{STAR} in central Au-Au collisions.}
\label{fig:energydep}
\end{figure}

The ALICE results for the most central Pb-Pb collisions are compared to NA49 \cite{NA49} and STAR \cite{STAR} data in 
Fig.~\ref{fig:energydep}. 
It should be noted here that the detector acceptances, momentum ranges and primary particle selection criteria are 
slightly different. The ALICE data indicate positive results for \nudyn in all three cases and are consistent with Poissonian 
expectation within 2$\sigma$.

\FloatBarrier
\section{Conclusions}

In summary, dynamical fluctuations of identified particle ratios were measured as function of centrality 
in Pb-Pb collisions at \snn=2.76~TeV with ALICE. The identity method, which is planned to be used in further event-by-event analyses in ALICE, 
was applied to ALICE data successfully for the first time. 
The results for \mbox{\nudynkapr} and \mbox{\nudynpika} are in qualitative agreement with models. 
However, the negative values of \mbox{\nudynpipr} observed in most peripheral collisions, which indicate a correlation 
between pions and protons, are not reproduced by the models. 
In all three cases, ALICE results in the most central events indicate positive results consistent with Poissonian 
expectation within 2$\sigma$ which agree with the extrapolations based on the data at lower energies from CERN-SPS and RHIC.
As a next step, a more differential analysis of \nudyn in terms of rapidity, \pt, and charge sign dependence 
as well as a study in p-p collisions is presently being prepared.

\FloatBarrier
\bibliographystyle{elsarticle-num}

\end{document}